\documentclass[twocolumn,trackchanges,twocolappendix]{aastex63}
\usepackage{ragged2e}
\usepackage{amsmath}
\usepackage[nameinlink]{cleveref}

\usepackage[]{natbib}
\setcitestyle{comma}
\usepackage{comment}
\usepackage{enumitem}

\usepackage[utf8]{inputenc}
\usepackage[T1]{fontenc}

\makeatletter
\newcommand*{\linktocite}[2]{%
  \hyper@natlinkstart{#1}#2\hyper@natlinkend}
\makeatother

\newcommand{\Eq}[1]{Equation\,(\ref{#1})}

\newcommand{\Sec}[1]{Section~\ref{#1}}

\newcommand{\Fig}[1]{Figure~\ref{#1}}

\def\K{\; {\rm K}}

\shorttitle{Tropical cyclones on tidally locked rocky planets}
\shortauthors{Garcia, Smith, Chavas, Komacek}
\graphicspath{{./}{figs/}}

\begin{document}
\turnoffedit

\title{Tropical {cyclones} on tidally locked rocky planets: Dependence on rotation period}

\author{Valeria Garcia}
\affiliation{Department of Atmospheric Sciences, University of Washington, Seattle, WA 98195, USA}
\affiliation{Department of Earth, Atmospheric, and Planetary Sciences, Purdue University, West Lafayette, IN 47907, USA}

\author{Cole M. Smith}
\affiliation{Department of Astronomy, University of Maryland, College Park, MD 20742, USA}

\author[0000-0001-9172-8328]{Daniel R. Chavas}
\altaffiliation{D.R.C. and T.D.K. contributed equally to this work.}
\affiliation{Department of Earth, Atmospheric, and Planetary Sciences, Purdue University, West Lafayette, IN 47907, USA}

\correspondingauthor{Thaddeus D. Komacek}
\email{tkomacek@umd.edu}
\author[0000-0002-9258-5311]{Thaddeus D. Komacek}
\altaffiliation{D.R.C. and T.D.K. contributed equally to this work.}
\affiliation{Department of Astronomy, University of Maryland, College Park, MD 20742, USA}




\begin{abstract}
Tropical cyclones occur over the Earth's tropical oceans, with characteristic genesis regions and tracks tied to the warm ocean surface that provides energy to sustain these storms. The study of tropical cyclogenesis and evolution on Earth has led to the development of environmental favorability metrics that predict the strength of potential storms from the local background climate state. Simulations of the gamut of transiting terrestrial exoplanets orbiting late-type stars may offer a test of this Earth-based understanding of tropical cyclogenesis. Previous work has demonstrated that tropical cyclones are likely to form on tidally locked terrestrial exoplanets with intermediate rotation periods of $\sim 8-10~\mathrm{days}$.
In this study, we test these expectations using ExoCAM simulations with both a sufficient horizontal resolution of 0.47$^\circ$ x 0.63$^\circ$ required to permit tropical cyclogenesis along with a thermodynamically active slab ocean. We conduct simulations of tidally locked and ocean-covered Earth-sized planets orbiting late-type M dwarf stars with varying rotation periods from 4-16 days in order to cross the predicted maximum in tropical cyclogenesis. We track tropical cyclones that form in each simulation and assess their {location of maximum wind}, evolution, and maximum wind speeds. We compare the resulting tropical {cyclone} locations and strengths to predictions based on environmental favorability metrics, finding good agreement between the Earth-based metrics and our simulated storms with a local maximum in both tropical cyclone frequency and intensity at a rotation period of 8 days. Our results suggest that {environmental favorability metrics used for tropical cyclones on Earth may also be applicable to temperate tidally locked Earth-sized rocky exoplanets with abundant surface liquid water}. 
\end{abstract}



\section{Introduction}
Tropical cyclones are one of the most impactful weather phenomena on Earth, characterized by fast winds 
and heavy rainfall \citep{Emanuel:2003aa}. Tropical cyclones act as heat engines, powered by heat input from the warm ocean surface and dissipated via friction in the boundary layer between the atmosphere and surface \citep{Emanuel:2003aa,Chavas:2017aa}. On Earth, tropical cyclones may have large-scale impacts on the mean climate, most notably by drying surrounding regions \edit1{\citep{Schenkel:2015ab}} and thereby enabling dry regions to more efficiently radiatively cool to space \citep{1995JAtS...52.1784P,wing2019self}. 
Tropical cyclones are also expected to form on tidally locked terrestrial exoplanets \citep{Komacek:2020ad,YanYang:2020aa} and may have potential consequences for observable properties, especially the amplitude of water vapor features in transmission spectra \citep{YanYang:2020aa} as well as their time-variability \citep{2021FrASS...8..134S,May:2021aa,2023ApJ...942L...4R}. 

Existing theories for the formation and evolution of tropical cyclones have identified a broad set of metrics for the environmental favorability required for tropical cyclogenesis to occur. These include the maximum potential intensity 
\citep{Bister:2002aa}, genesis potential index (GPI) 
\citep{Emanuel:2010aa}, ventilation index 
\citep{Tang:2012aa}, and ventilation-reduced potential intensity 
\citep{Chavas:2017aa}, along with more general quantities including absolute vorticity \citep{Hoogewind:2020aa} and climate properties including sea surface temperature and humidity \citep{Camargo:2014aa}. These favorability metrics, especially the combination of absolute vorticity, maximum potential intensity, and ventilation index, can successfully match the observed regions of tropical cyclogenesis on Earth \citep{Hoogewind:2020aa}. Recent work has calculated these favorability metrics from low-resolution general circulation model (GCM) simulations of tidally locked rocky exoplanets, finding that tropical cyclones should be unlikely to occur on slowly rotating exoplanets but peak in environmental favorability at intermediate rotation periods of $\sim 8-10$ days \citep{Bin:2018aa,Komacek:2020ad}. 

Idealized experiments with a sufficient spatial resolution to permit tropical cyclones have investigated the dependence of cyclogenesis and intensity on key climate controls including moisture availability and rotation rate \citep{ReedChavas:2015,Merlis:2016aa,Chavas:2019aa,Cronin:2019aa}. This approach can be extended to simulate the potential for tropical cyclogenesis in the atmospheres of terrestrial exoplanets. Notably, \cite{YanYang:2020aa} conducted $\approx 50~\mathrm{km}$ horizontal resolution GCMs of tidally locked terrestrial exoplanets with prescribed sea surface temperature (SST) distributions. They found that, as expected from environmental favorability metrics, tropical cyclones may be common on terrestrial exoplanets with plentiful surface oceans. However, the prescribed day-night SST gradients in their simulations that produced tropical cyclones were smaller than might be expected from simulations with an interactive slab ocean. As a result, there is a need to further test our largely Earth-based understanding of tropical cyclogenesis using models with a thermodynamic ocean that cover a broader range of planetary parameters.

In this manuscript, we improve on \cite{YanYang:2020aa} by conducting {storm-resolving} GCM simulations at an equivalent horizontal resolution but with a thermodynamically active slab ocean and thermodynamic sea ice scheme. We conduct GCMs of tidally locked planets with varying rotation periods in order to test the expectation from Earth-based environmental favorability metrics that tropical cyclone activity on tidally locked terrestrial exoplanets is most pronounced at intermediate rotation periods of $\sim 8-10$ Earth days. Specifically, we test the results of \cite{Komacek:2020ad} by conducting otherwise equivalent GCMs but with sufficient horizontal resolution to permit tropical cyclogenesis. This work is organized as follows. We describe our GCM setup and tropical cyclone tracking algorithm in \Sec{sec:methods}. In \Sec{sec:results}, we present results for how tropical cyclone location and intensity varies with rotation period and then compare our numerical simulations with environmental favorability metrics. We discuss implications of our results as well as limitations of our model setup in \Sec{sec:disc}. Finally, we summarize key points in \Sec{sec:conc}.

\section{Methods}
\label{sec:methods}
\subsection{GCM simulations}
In order to simulate the atmospheric circulation of tidally locked rocky exoplanets orbiting late-type M dwarf stars, we use the ExoCAM\footnote{ \url{https://github.com/storyofthewolf/ExoCAM}} GCM \citep{Wolf:2022aa}. ExoCAM has a broad heritage in this parameter regime \citep{kopparapu2017,Wolf:2017aa,Haqq2018,Komacek:2019aa-terrestrial,Yang:2019aa,Suissa:2020aa,2020ApJ...898..156W,Wolf:2020aa,Wolf:2022aa} and has participated in multiple broad 3D GCM inter-comparison studies for the climates of potentially habitable rocky exoplanets orbiting M dwarf stars, including the TRAPPIST-1 Habitable Atmosphere Intercomparison (THAI) project \citep{2019ApJ...875...46Y,2022PSJ.....3..213F,2022PSJ.....3..212S,2022PSJ.....3..211T}. ExoCAM itself is a branch of the Community Atmosphere Model Version 4 (CAM4), and it is coupled to ExoRT\footnote{\url{https://github.com/storyofthewolf/ExoRT}}, a flexible correlated-k radiative transfer routine that is applicable to exoplanet atmospheres near the inner edge of the habitable zone. ExoRT opacities are sourced both from HITRAN and the HELIOS-K opacity calculator \citep{2021ApJS..253...30G,2022JQSRT.27707949G}. 

In this work, we use ExoCAM and ExoRT to simulate the circulation of tidally locked rocky exoplanets solely as a function of rotation period. Specifically, we conduct branch simulations from the ExoCAM GCMs presented in \cite{Komacek:2020ad} with sidereal rotation periods of 4, 8, and 16 days\footnote{Throughout this work, ``day'' refers to an Earth day.}. To do so, we interpolate the equilibrated GCM output from cases with a horizontal resolution of $4^\circ \times 5^\circ$ in \cite{Komacek:2020ad} to a 
horizontal resolution of $0.47^\circ \times 0.63^\circ$ in latitude and longitude, as in \cite{Komacek:2019ab}. This horizontal resolution has previously been demonstrated to permit tropical cyclones in GCM simulations of tidally locked terrestrial planets with fixed sea surface temperature \citep{YanYang:2020aa}. All simulations use planetary parameters (besides rotation period/rate) that are identical to the Earth, including a radius of $6.37122 \times 10^6~\mathrm{m}$, surface gravity of $9.80616~\mathrm{m}~\mathrm{s}^{-2}$, and stellar constant of $1360~\mathrm{W}~\mathrm{m}^{-2}$. We set obliquity and eccentricity to zero, consistent with the assumption of tidal locking. Note that these idealized simulations do not consistently vary the incident stellar flux and rotation period (according to Kepler's 3rd law, \citealp{kopparapu2017}) in order to isolate the effect of rotation period on tropical cyclogenesis. The atmospheric composition is set to have 1 bar of background N$_2$, along with water vapor set by Clausius-Clapeyron assuming an aquaplanet surface. All simulations use an incident stellar spectrum of a late-type M dwarf star with $T_\mathrm{eff} = 2600~\mathrm{K}$ \citep{Allard:2007aa}. We further include a 50 m slab ocean\footnote{\cite{Komacek:2020ad} previously demonstrated that using a shallow 1 m slab ocean depth does not affect predictions for tropical cyclogenesis, and thus in this work we only study cases with a thick slab ocean and resulting high surface heat capacity.} and the CESM thermodynamic sea ice scheme \citep{Bitz:seaice2001,Bitz:2012aa}, while notably neglecting sea ice drift \citep{Yang:2020aa}. 

The simulations presented in this work use a dynamical timestep of 30 minutes and a radiative timestep of 1 hour. Each simulation has 40 vertical layers, spaced as in \cite{Komacek:2020ad}. We use 68 spectral intervals in the ExoRT correlated-k scheme (version src.cam.n68equiv), ranging from 0.238 $\mu m$ - $\infty$. From the equilibrated initial state, we conduct simulations for 18 months. We only process the final 6 months of the simulation output in our analysis while the first 12 months are discarded as an initial re-adjustment period. The analyses presented in this work focus on the 6-hourly climatology used for studying both the structure of and tracking tropical cyclones. 

\subsection{Cyclone tracking}
In order to track candidate tropical cyclones in our ExoCAM simulations, we process our GCM ouptut with TempestExtremes \citep{Ullrich:2021aa}. TempestExtremes is a generalized feature tracker that tracks tropical cyclones by defining features using closed contour criteria. Our tropical cyclone tracking process with TempestExtremes consists of two discrete steps. First, we search for potential locations of tropical cyclones in each of our 6-hourly output timesteps using the DetectNodes routine in TempestExtremes. Specifically, we search for closed contours around minima in surface pressure with a surface pressure decrement of at least {2 mbar} from the environment. \edit1{These minima must also have a geopotential contrast of $-58.8~\mathrm{m}^{2}~\mathrm{s}^{-2}$ between 300 and 500 mbar in order to ensure that they are warm core storms \citep{Zarzycki:2017aa}. This second criterion is important to distinguish tropical cyclones from other low pressure disturbances.} We allow maximum closed contour widths of $6^\circ$, $12^\circ$, and $24^\circ$ in cases with rotation periods of 4 days, 8 days, and 16 days, respectively. We increase the allowed closed contour width with rotation period in order to account for the expected increase in storm size  \citep{Chavas:2019aa,Lu:2022aa}. We then ``stitch'' these detected low-pressure regions in time to form tropical cyclone trajectories with the StitchNodes routine in TempestExtremes. We choose a maximum range of $4^\circ$ for tropical cyclone stitching over the 6 hours between output timesteps, a minimum storm length of 24 hours, and a maximum gap between consecutive stitched points of 12 hours. 
\section{Results}
\label{sec:results}
\subsection{Tropical cyclones with varying rotation period}
\label{sec:results1}
\begin{figure*}
    \centering
    \includegraphics[width=1\textwidth]{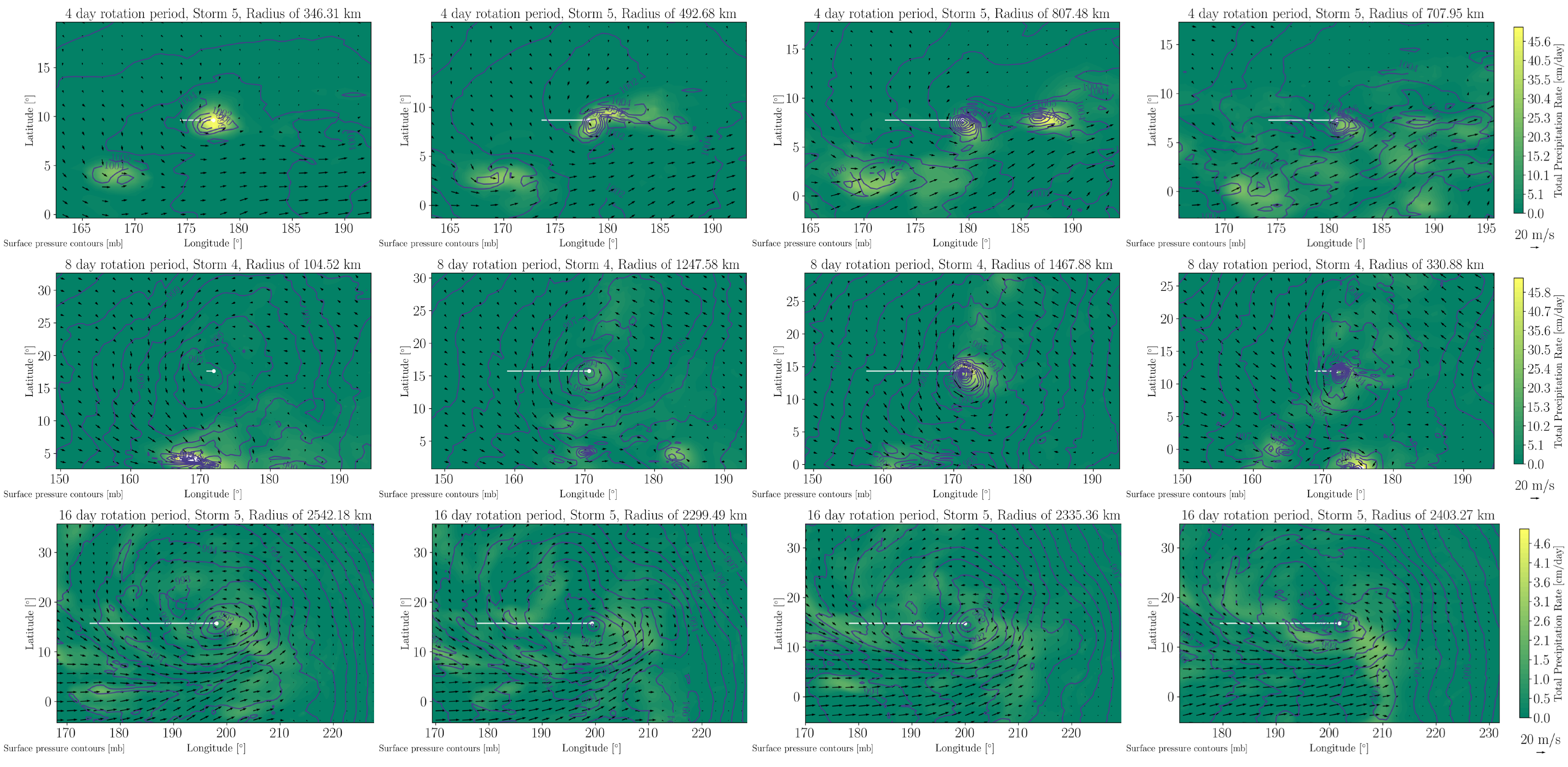}
    \caption{Snapshots of individual storms from cases with a rotation period of 4 days (top), 8 days (middle), and 16 days (bottom). The total precipitation rate is plotted in filled contours, the surface pressure (in {mbars}) is plotted in the blue contours, and near-surface wind speeds (in m/s) are plotted in quivers. \edit1{Snapshots are taken in 6-hourly intervals, and each set of snapshots cross the time of maximum intensification of the storm}. The white bar demarcates the characteristic radius of the storm, which is further listed in the figure title. Note that the substellar longitude is $180^\circ$, and that the scaling of the longitude and latitude range and surface pressure changes between panels.}
    \label{fig:case}
\end{figure*}

Before studying the properties of the full sample of tracked storms from our simulations, we {identified} the five storms with the longest lifetime from each of the three simulations with varying rotation period, \edit1{labeling them by number in order of decreasing tracked point count (i.e., storm length)}. The horizontal structure, including surface pressure, surface winds, and precipitation, in one \edit1{of the five} storm case studies \edit1{over the course of the 24 hours crossing the timing of maximum intensification} for each simulation is shown in \Fig{fig:case}. 
 \\
\indent \edit1{As expected,} the tracked storms are each cyclonic (with anti-clockwise winds if in the northern hemisphere and clockwise winds in the southern hemisphere) and associated with a pressure decrease toward the center. The case study in the simulation with a rotation period of 8 days has the greatest pressure gradient from center of storm outward along with the strongest winds 
as expected from geostrophic balance. Note that some storms in the case studies with a rotation period of 8 days are accompanied by a cyclone in the southern hemisphere (not shown) -- such dipolar cyclone, or ``modon'' \citep{Cho:2021wb} structure \edit1{has been found in previous tidally locked planet GCMs.}

The white bar in each panel of Figure \ref{fig:case} shows the characteristic radius of the storm, calculated as the radius where the azimuthal average of the winds decreases below 8 m/s. The characteristic radius increases with increasing rotation period in these specific case studies, as expected given the increase of the Rhines and inverse f-scales with rotation period \citep{Chavas:2019aa,Lu:2022aa}. The precipitation peaks around the pressure minimum in all cases. The storm intensity, measured by both the horizontal surface pressure minimum and wind speeds, is largest in the 8 day case. Additionally, the storms are more symmetric in the 8 day case than in the 4 and 16 day cases -- in the 4 day case, the strong wind shear due to the superrotating jet elongates the tropical cyclone, while in the 16 day case there is significant sub-structure in the storm at scales below the radius of maximum wind. Notably, in the 4 day case the wind quivers cross the center of the storm (rather than being in gradient wind balance), which demonstrates the influence of the background flow on the storm structure \edit1{of this weaker cyclone}. 

\begin{figure*}
    \centering
    \includegraphics[width=0.85\textwidth]{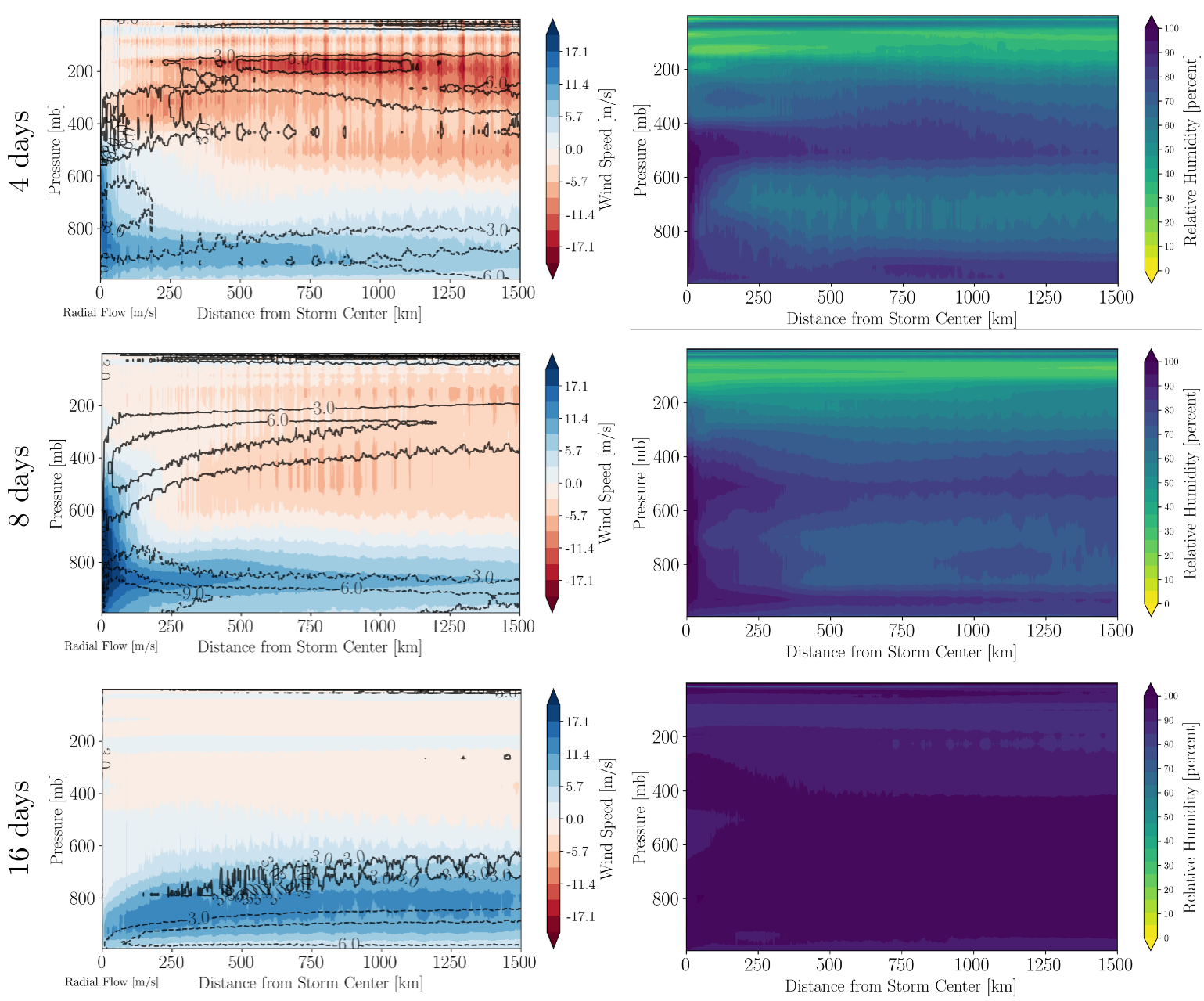}
    \caption{Storm-centered cross-sections for the same storms in Figure \ref{fig:case} from cases with a rotation period of 4 days (top), 8 days (middle), and 16 days (bottom). Each vertical level is averaged azimuthally in circles around the tracked storm center. The left-hand panels display the azimuthally averaged cyclonic wind speed (in $\mathrm{m}~\mathrm{s}^{-1}$, with positive values representing cyclonic flow and negative values representing anticyclonic flow), and the right-hand panels display the azimuthally averaged relative humidity (in $\%$). The contour levels on the left-hand panel display radial wind speeds in $\mathrm{m}~\mathrm{s}^{-1}$\edit1{, with inflow denoted by dashed contours and outflow by solid contours}. The vertical axis is the atmospheric pressure (in millibars) and the horizontal axis is the distance from the tracked storm center (in km). Each cross-section shown is a snapshot at the {time of maximum intensification} in Figure \ref{fig:case}.}
    \label{fig:casevert}
\end{figure*}

Figure \ref{fig:casevert} shows azimuthally averaged vertical cross sections of cyclonic and radial wind speeds as well as relative humidity from the same storms shown in Figure \ref{fig:case}. The vertical cross sections in Figure \ref{fig:casevert} are further shown from the same model output time as the horizontal profiles in Figure \ref{fig:case}. All azimuthal averages are performed on isobars, and the cyclonic wind magnitude is specifically the azimuthal storm-centered average of the rotational (i.e., non-divergent) component in the Helmholtz decomposition \citep{Hammond:2021aa}. 

We find that the center of each storm is associated with a tower of high relative humidity extending to the top of the troposphere, as expected from the previous tropical cyclone permitting simulations of \cite{YanYang:2020aa}. All storms have radial inflow near the surface and outflow at upper levels, as expected for tropical cyclones \citep{Emanuel:2003aa}. We also find a peak in the cyclonic wind speeds in the deep atmosphere adjacent to the center of the storm (i.e., an eyewall-like structure) in the case study with a rotation period of 16 days, as well as in other case studies with a rotation period of 4 and 8 days (not shown). In the fast-rotating cases with rotation periods of 4 and 8 days, we find that the presence of a superrotating jet at low pressures 
ventilates the top of the storm (\edit1{note the band of small relative humidity at low pressures}), limiting the height of the tropical cyclone. The 8 day case shows the best correspondence of horizontal and vertical structure to both expectations of tropical cyclones for Earth as well as the previous tidally locked exoplanet GCMs of \cite{YanYang:2020aa}. Overall, all of our case studies with varying rotation period are tropical cyclone-like, consisting of a surface-based vortex that is strongest at small radii, with inflow at low levels and outflow at upper levels, a nearly-saturated convecting column at the center, and (in some cases) an eye-like feature at the center. 

\begin{figure*}
    \centering
    \includegraphics[width=0.9\textwidth]{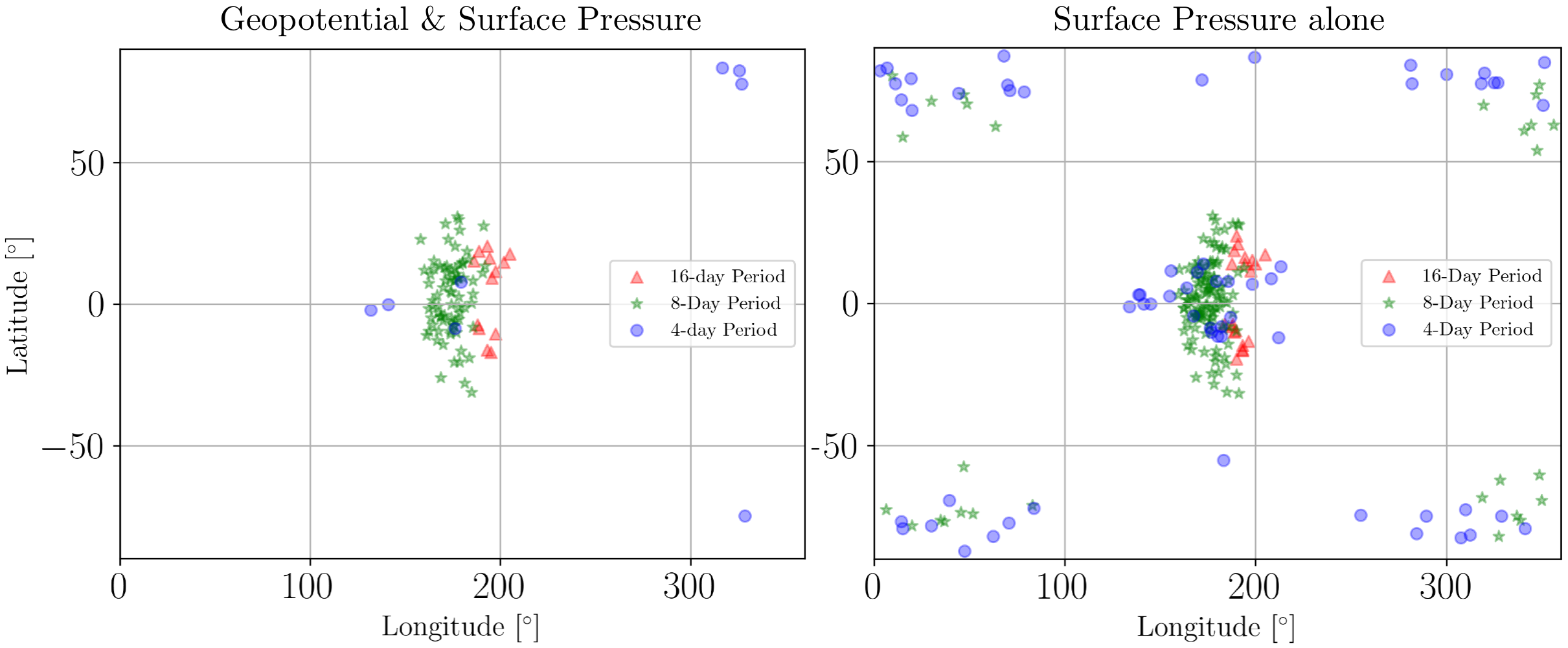}
    \caption{Location of where each tracked storm has its maximum total horizontal wind speed. \edit1{The left-hand plot shows the results when including both surface pressure and geopotential criterion in the tracking, while the right-hand side shows the results from tracking that only considers surface pressure (and hence does not distinguish between warm and cold core storms).} 16 day period cases are shown as red triangles, 8 day cases as green stars, and 4 day cases as blue circles. \edit1{Storms near the polar regions in the surface pressure alone tracking} are polar cyclones associated with the Rossby gyres of the \edit1{planetary-scale} Matsuno-Gill pattern, most commonly tracked in faster-rotating cases. All near-equatorial storms are found on the dayside, concentrated around the substellar point.}
    \label{fig:loc}
\end{figure*}

To provide context, the location at which each tracked storm from the three simulations had its strongest horizontal wind is shown in \Fig{fig:loc}.
\edit1{We study the location of peak intensity rather than genesis because the location of peak intensity best corresponds to the distribution of where tropical cyclones can be supported by the environment. We also show tracking both including and excluding the geopotential criterion in order to display the importance of including geopotential in tracking storms on tidally locked rocky planets.}  There are two classes of tracked storms that are immediately distinguishable \edit1{by comparing tracking with and without the geopotential criterion}: tropical (warm-core) cyclones that form near the substellar point, and extratropical (cold-core) cyclones that form near the polar regions. We have visually verified that the extratropical cyclones are cold-core and associated with the Rossby gyres of the planetary-scale Matsuno-Gill pattern \citep{Showman_2013_terrestrial_review,Pierrehumbert:2019aa}.

As evident in \Fig{fig:loc}, all of the warm-core tropical cyclones in our three GCMs occur on the dayside, concentrated near the substellar point. The patterns of the tracked points differ between the cases with varying rotation period, with the near-equatorial storms in the 4-day case concentrated {closer to} the equator, those in the 8 day case in a crescent that extends north and south from the substellar point, and those in the 16 case concentrated in two regions northeast and southeast of the substellar point. As we discuss in \Sec{sec:envfav}, the locations of the maximum wind for the tropical cyclones are controlled by the background climate, and thus can be predicted from large-scale environmental favorability metrics. \edit1{As we will show, the locations of storms move further from the equator in more slowly rotating cases due to the constraint of sufficient absolute vorticity, which is smaller near the equator due to a combination of slower rotation and weaker winds.}

\subsection{Comparison with environmental favorability metrics}
\label{sec:envfav}
We compare our results to two separate environmental favorability metrics: the ventilation-reduced maximum potential intensity (a metric which incorporates the ventilation index to calculate a refined maximum potential intensity) and absolute vorticity. The ventilation-reduced maximum potential intensity provides the maximum achievable wind speed for a tropical cyclone in the presence of ventilation by dry air from its surroundings. The normalized ventilation-reduced maximum potential intensity can be written in a simpler form than \cite{Komacek:2020ad} following some mathematical reduction:
\begin{equation}\label{eq:PIvent1}
    u_\mathrm{p,VI} = \left(x+\frac{1}{3x}\right)u_\mathrm{p} \mathrm{,}
\end{equation}
where
\begin{equation}\label{eq:PIvent2}
    x = \frac{1}{\sqrt{3}}\left[\sqrt{\left(\left(\frac{VI}{VI_\mathrm{thresh}}\right)^2-1\right)}-\left(\frac{VI}{VI_\mathrm{thresh}}\right)\right]^\frac{1}{3} \mathrm{,}
\end{equation}
and $u_\mathrm{p}$ is the maximum potential intensity \citep{Bister:2002aa}.

Equations (\ref{eq:PIvent1}) and (\ref{eq:PIvent2}) are derived from the equilibrium solution for normalized intensity in the presence of mid-level ventilation \citep{Chavas:2017aa}. Here VI is the ventilation index calculated as in \cite{Tang:2012aa}.
In Equation (\ref{eq:PIvent2}), $VI_\mathrm{thresh}$ is a threshold value of VI representing the maximum value of VI where $u_\mathrm{p,VI}>0$ and no storm can be sustained. We set $VI_\mathrm{thresh}=0.145$ which was estimated based on Earth in \cite{Hoogewind:2020aa} and used in \cite{Komacek:2020ad}.
This solution can have both real and imaginary components; only the real component is physical and is used as the solution.

\begin{figure*}
    \centering
    \includegraphics[width=0.99\textwidth]{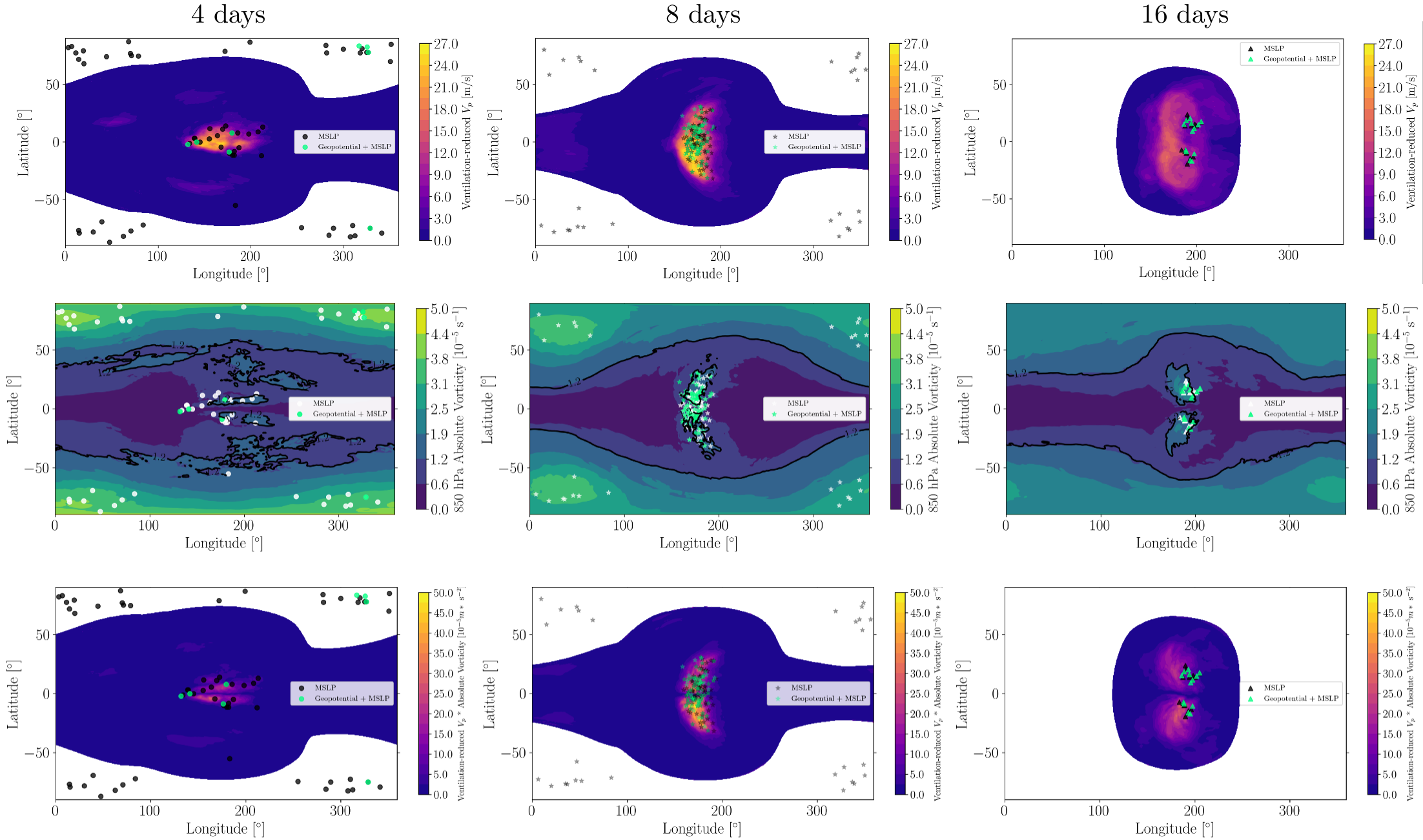}
    \caption{\edit1{Comparison of environmental favorability metrics from the mean climate state to our tracked cyclone locations.} Top row: 6-month mean ventilation reduced maximum potential intensity (filled contours) for each case with varying rotation period, with locations of storms from Figure \ref{fig:loc} over-plotted, \edit1{both from tracking with just surface pressure (MSLP) and both surface pressure and geopotential}. Middle: Same as the top row, but for absolute vorticity. Bottom: Same as above, but for the product of ventilation-reduced maximum potential intensity and absolute vorticity. We find that \edit1{warm-core} tracked storm locations are generally found near maxima in ventilation-reduced maximum potential intensity and the product of ventilation-reduced maximum potential intensity and absolute vorticity, with especially good agreement in the case with a rotation period of 8 days.}
    \label{fig:virvp_absvort}
\end{figure*}

The top row of \Fig{fig:virvp_absvort} compares tropical cyclone locations both using surface pressure (labeled MSLP, \footnote{Note that mean surface pressure and mean sea level pressure are equivalent in these aquaplanet simulations.}) and surface pressure and geopotential together from Figure \ref{fig:loc} to the six-month mean of $u_\mathrm{p,VI}$ for each of the three cases. \edit1{In this and all comparisons to environmental favorability metrics, we take the tropical cyclone location to be the location of peak intensity -- i.e., where the storm has a maximum in its tracked 850 hPa horizontal wind speed. As mentioned in Section \ref{sec:results1}, this is because the location of maximum intensity will correspond to the point at which the storm is best supported by the environment.}
We find that the simulated tropical cyclones cluster near the maxima of $u_\mathrm{p,VI}$, with especially good agreement in the 8 day rotation period case. However, this metric alone does not explain the eastward shift of storms from the substellar point in the 4 and 16 day cases. 

In order to determine the extent to which multiple metrics can explain tropical cyclone locations, we also compare our simulated tropical cyclone locations with the \edit1{six-month mean of the} 850 mbar absolute vorticity $\eta$, which is simply the sum of the relative vorticity $\xi$ and planetary vorticity $f$ at that level: 
\begin{equation}
\label{eq:vorticity}
\eta = \xi + f = \hat{k} \cdot (\nabla \times {\bf u}) + 2 \Omega~\mathrm{sin}(\phi). 
\end{equation}
In \Eq{eq:vorticity}, $\Omega$ is the rotation rate, $\phi$ is the latitude, $\hat{k}$ is the unit vector in the vertical direction, and ${\bf u}$ is the vector wind speed. The middle row of \Fig{fig:virvp_absvort} compares the locations of simulated tropical cyclones with $\eta$. We find that the locations of storms in the 16 day case are well-correlated with absolute vorticity maxima, and in the 4 day case the maxima of absolute vorticity poleward of the substellar point better match with the tracked cyclone locations. However, there is a large clustering of storms near the substellar point in the 8 day case that is not well explained by absolute vorticity alone, as it peaks at higher latitudes. 

We postulate that a simple product of absolute vorticity and ventilation-reduced maximum potential intensity, which is similar to existing GPIs as shown in \cite{Komacek:2020ad}, may qualitatively describe the locations of simulated tropical cyclones. The bottom row of \Fig{fig:virvp_absvort} compares storm locations to the product $\eta \times u_\mathrm{p,VI}$. We find that this combined metric provides the best overall match to the locations of tropical storms, especially in the 4 and 8 day cases. In the 16 day case, absolute vorticity alone is a better predictor of storm location than the combination of absolute vorticity and ventilation-reduced maximum potential intensity, as the surface area with sufficient absolute vorticity in this region is relatively small and hence absolute vorticity is the primary bottleneck to genesis. Overall, we find that the case with an 8-day rotation period has the best alignment between high ventilation-reduced potential intensity and high absolute vorticity, and hence the most coherent region of high favorability \edit1{in the long-term mean of environmental properties} that aligns remarkably well with actual tracked tropical cyclone activity.

\subsection{Tropical cyclone properties}
\begin{figure}
    \includegraphics[width=0.44\textwidth]{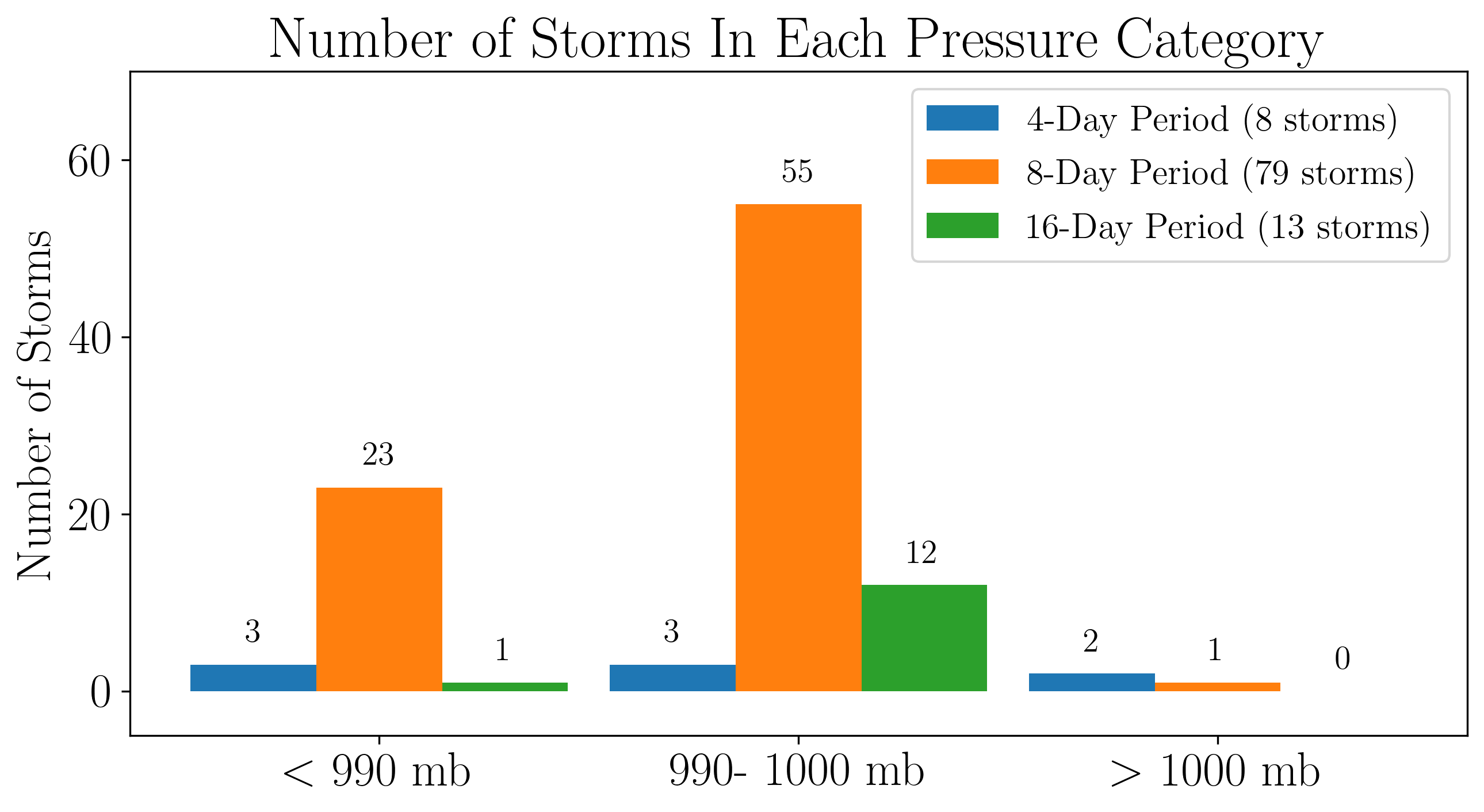}
    \includegraphics[width=0.48\textwidth]{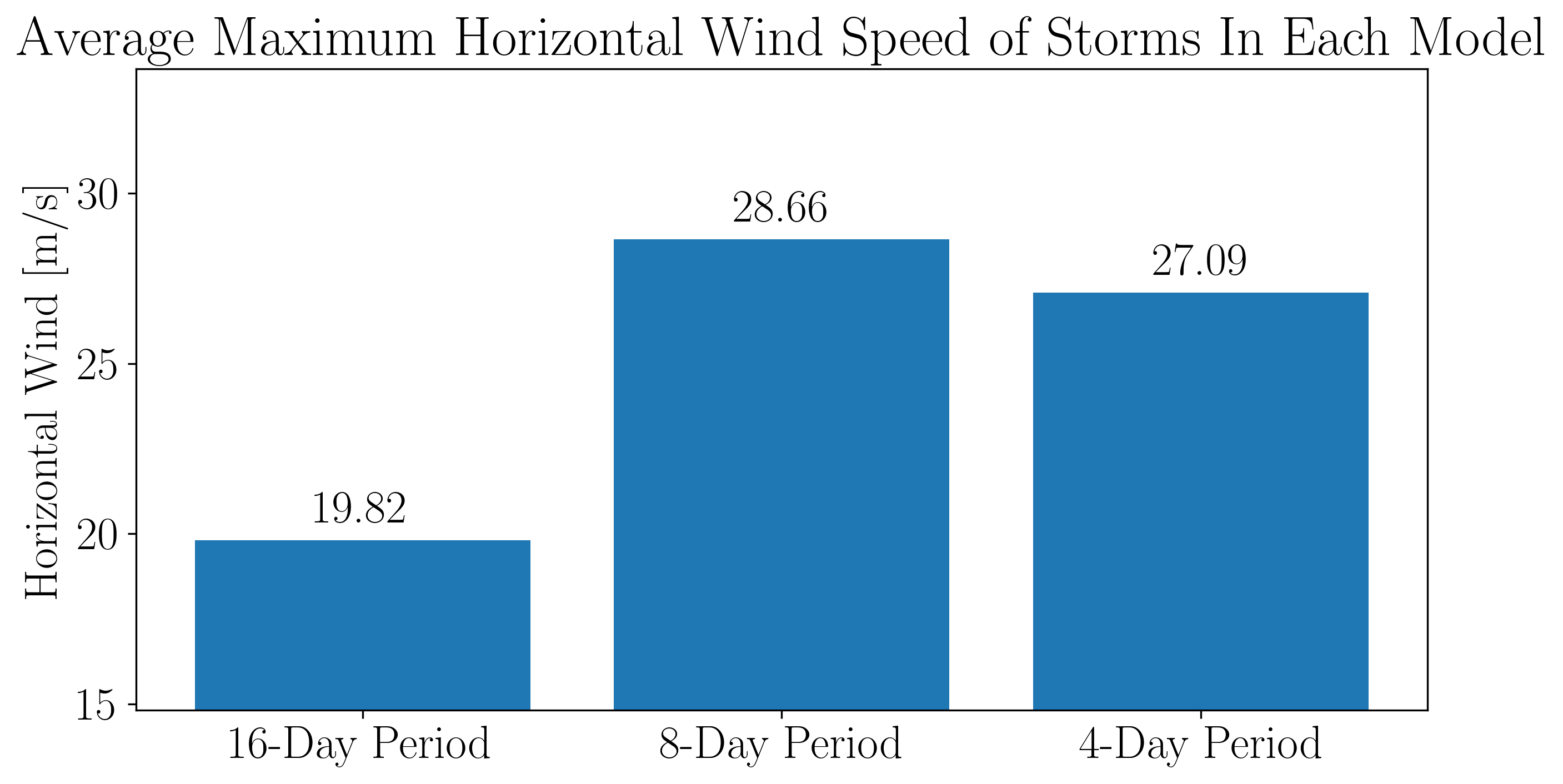}
    \includegraphics[width=0.44\textwidth]{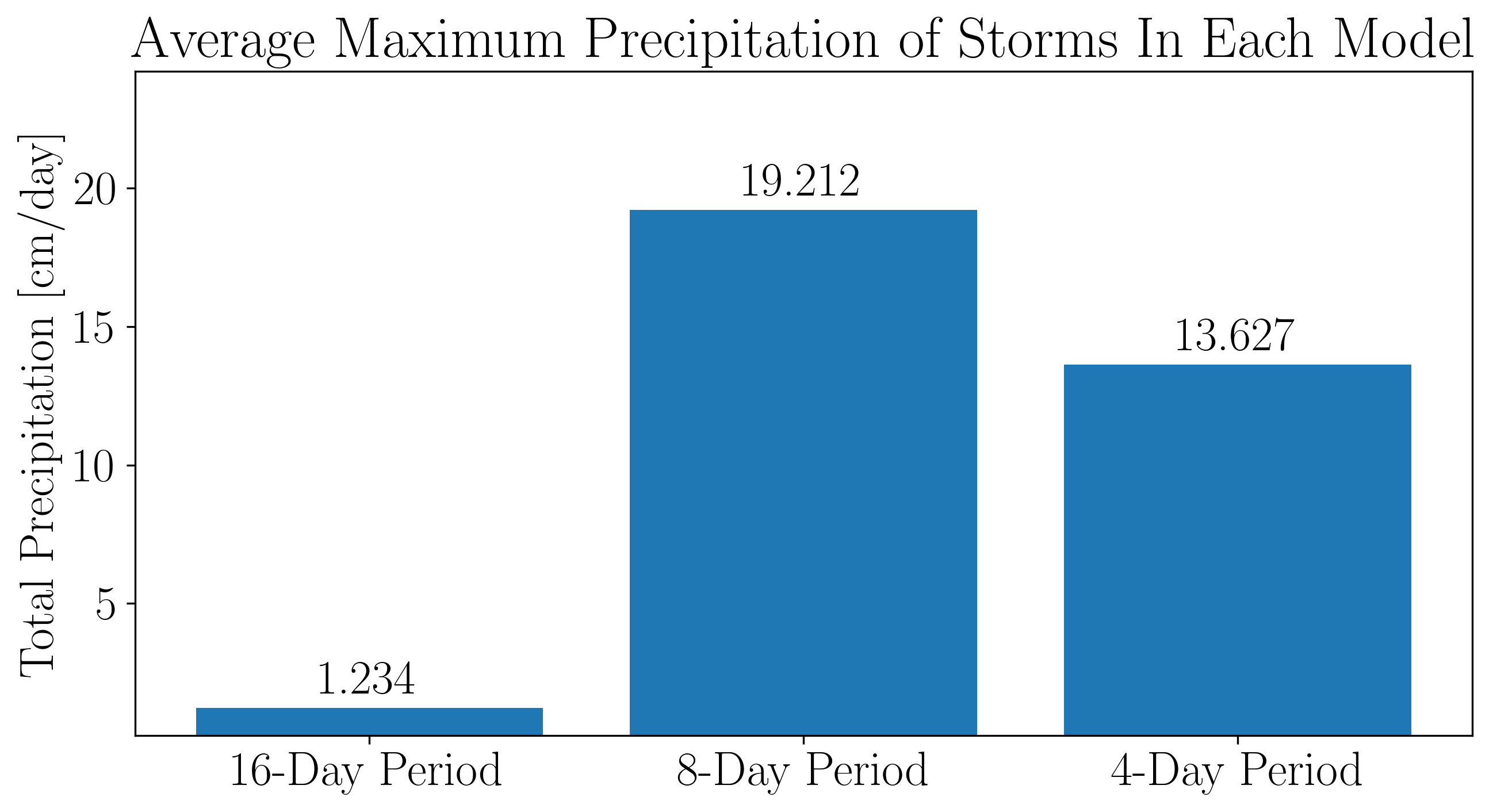}
    \caption{Summary histograms for the number distribution of minimum central tracked storm {surface} pressure with varying rotation period (top), and the average of the maximum wind speeds \edit1{from each storm} (middle) and \edit1{the average of the maximum} precipitation \edit1{from each storm} (bottom). \edit1{Results are shown for} all \edit1{tracked} storms \edit1{in our three cases} with varying rotation periods. We find that the 8 day case has the most storms overall and the most storms with hurricane-level {surface pressure}, resulting in the highest average maximum horizontal wind speeds. The 4 day case has the highest average of the maximum precipitation in the tracked storms, potentially because this simulation has the warmest sea surface temperatures.}
    \label{fig:stats}
\end{figure}

Finally, we summarize the intensity of storms with varying rotation period with {frequentist} statistics. \Fig{fig:stats} compares summary statistics for storms in each of our three simulations, {focusing only on the tracking with both geopotential and surface pressure constraints}\footnote{Note that far more storms are tracked when considering surface pressure alone. In the 4 day case, 62 storms are tracked with just surface pressure, while only 8 are found with both surface pressure and geopotential. Similarly, for the 8 day case 151 storms are tracked with just surface pressure and 79 with both surface pressure and geopotential, and in the 16 day case 20 storms are tracked with just surface pressure constraints and 13 storms are tracked with both surface pressure and geopotential.}.
The top row counts the number of storms with sufficient minimum {surface} pressure (\edit1{here equivalent to mean sea level pressure,} MSLP) \edit1{at any given point in their evolution} to be classified as hurricanes (MSLP $< 990~\mathrm{mbars}$) according to the revised Saffir–Simpson hurricane wind scale with modified MSLP \edit1{(i.e., surface pressure)} criteria of \cite{Klotzbach:2020aa}, along with storms of intermediate intensity ($990~\mathrm{mbars} <$ MSLP $< 1000~\mathrm{mbars}$) and low intensity (MSLP $> 1000~\mathrm{mbars}$). We find that over the six-month {output time} of the numerical experiments, the 8 day rotation period case has the most tracked systems with {surface pressures} corresponding to hurricanes, as well as the most tracked storms with intermediate intensity. \edit1{The 16 day case has the second-most storms with intermediate intensity, but zero hurricane-strength surface pressure decrements tracked over six months. Few warm-core storms form in the 4 day case, with only 8 total storms tracked, but with a roughly equal distribution of intensities in our coarse pressure bins.}

The middle row of \Fig{fig:stats} shows a measure of the average peak intensity, measured as the average of the maximum wind speeds at 850 mbars within $4^\circ$ of the tracked cyclone center in TempestExtremes. Additionally, the bottom row of \Fig{fig:stats} shows the average peak precipitation within $4^\circ$ of all storms for each case with varying rotation period. We find that the 8-day case has the greatest average storm wind speed, with the 16-day case having the weakest overall winds. The average maximum precipitation in the 8-day case is comparable to that of strong tropical cyclones on Earth, together with the wind speeds demonstrating that the bulk properties of tropical cyclones on exoplanets may not be dissimilar to those on Earth. Lastly, the 8-day case has the greatest average maximum precipitation, with the storms in the 16-day case conversely being associated with relatively little precipitation. The decrease in precipitation at longer rotation periods may be linked to the cooler sea surface temperatures in the slower-rotating cases due to the prevalence of dayside clouds (\citealp{Yang:2014}, \edit1{for results from GCM simulations with the same rotation period parameter sweep as conducted here, see Figure 9 of \citealp{Komacek:2019aa-terrestrial})}, as previous fixed-SST aquaplanet simulations have found that tropical cyclone precipitation correlates with SST \citep{Stansfield:2021aa}.  

\section{Discussion}
\label{sec:disc}
\subsection{Comparison to previous work}
Our results generally agree well with the previous expectations of \cite{Bin:2018aa} and \cite{Komacek:2020ad} from applying environmental favorability metrics to low-resolution exoplanet GCMs. As expected from the calculation of GPI in the slowly-rotating models with a rotation period of 28 days in \cite{Bin:2018aa}, we find that tropical cyclone intensity and number decreases sharply in our case with a rotation period of 16 days relative to faster-rotating cases. As expected from the work of \cite{Komacek:2020ad} applying the ventilation-reduced maximum potential intensity and absolute vorticity, we find that the number and intensity of tropical cyclones both peak in our intermediate rotation case with a rotation period of 8 days. The key difference \edit1{in our methodology compared to \cite{Bin:2018aa} and \cite{Komacek:2020ad} is that the simulations presented here have sufficient horizontal model resolution to resolve and track the tropical cyclones. One notable difference} between our findings and previous work is that we find weak tropical cyclone-like features (in terms of structure) which form at a rotation period of 16 days -- meanwhile, both \cite{Bin:2018aa} and \cite{Komacek:2020ad} predict that tropical cyclogenesis is not favorable at such slow rotation periods. Future work with convection-permitting models is needed to investigate the structure of our tropical cyclone features in the slowly rotating regime in order to determine the interplay between substellar convection and cyclogenesis.

We also find broad agreement with the fixed SST simulations presented in \cite{YanYang:2020aa}, as they find that cyclogenesis is prevalent in all models with intermediate or rapid rotation. The key difference between this work and \cite{YanYang:2020aa} is that we find that tropical cyclones only occur on the dayside near the substellar point. Meanwhile, tropical cyclones in \cite{YanYang:2020aa} can persist in non-irradiated regions near the terminator, likely because the nightside is hotter due to smaller prescribed day-to-night SST contrasts. Meanwhile, in our simulations, the nightside has a large enough ventilation index that the ventilation-reduced maximum potential intensity is near-zero or undefined, as the sea surface is cooler than the overlying atmosphere and thus the net heat flux is downwards into the surface (Figure \ref{fig:virvp_absvort}). This difference between our results and \cite{YanYang:2020aa} may depend on background climate state and thus assumed atmospheric abundance of carbon dioxide or other greenhouse gases, assumed to be zero in this work. In general, we expect that tropical cyclones are more likely to persist on the nightside in hotter and moister climates, given that for temperate terrestrial planets day-to-night temperature contrasts decrease with increasing irradiation \citep{Haqq2018}.

\subsection{Utility of Earth-based tropical cyclogenesis favorability metrics}
We find that the Earth-based metrics of absolute vorticity, ventilation index, maximum potential intensity, and ventilation-reduced maximum potential intensity presented in \cite{Chavas:2017aa} and \cite{Hoogewind:2020aa} accurately predict the locations of tropical cyclones on tropical cyclone-permitting models of tidally locked terrestrial exoplanets. Notably, the ventilation-reduced maximum potential intensity \citep{Chavas:2017aa,Komacek:2020ad} is a novel metric that has been applied to exoplanet simulations before being verified for Earth. Additional work applying the ventilation-reduced maximum potential intensity back to Earth itself finds good agreement with observed tropical cyclone genesis locations (Chavas et al., in prep.). Continued exoplanet-specific and idealized simulations of tropical cyclones may provide further insight on the properties that govern tropical cyclone behavior on Earth. 

\subsection{Limitations and future work}
Though we improve on previous work by conducting tropical cyclone permitting simulations of tidally locked exoplanets with thermodynamic ocean and sea ice schemes, our model setup is still idealized in many respects. Notably, we do not include a dynamic ocean with motions forced by wind stress from the overlying atmosphere, which has been shown to result in a characteristic eastward offset in the sea surface temperature maximum (known as a ``lobster'' planet, \citealp{Hu:2014aa}). We also assume an aquaplanet and thus do not include any continents in our simulations, which have been shown to potentially have a large effect on moisture availability and surface enthalpy fluxes \citep{Lewis:2018aa,Salazar:2020aa,Macondald:2022aa}. We anticipate that including a dynamic ocean and/or continents may shift the region of tropical cyclone favorability toward the regions near the equator with the warmest sea surface temperatures. In the case of a large sub-stellar continent, it is possible that the low heat capacity (sensible plus latent) \edit1{along with the limited availability of moisture and lack of oceanic latent heat fluxes} of the surface may prevent tropical cyclogenesis. Future work is needed to simulate the impact of dynamic oceans and continent distribution on the prevalence of tropical cyclones on tidally locked terrestrial planets. \edit1{In addition, further simulations are needed to assess the potential for tropical cyclogenesis on specific nearby temperate rocky exoplanets, including Proxima Centuari b, TRAPPIST-1e, and LP 890-9c.} 

Our global GCMs are near the low end of the level of resolution required to {resolve the basic structure of} tropical cyclones, and in global models the resolution may impact the frequency and intensity of tropical cyclones \citep{Vecchi:2019}. In addition to being coarse in horizontal resolution, our simulations use the hydrostatic primitive equations and thus would not be convection-permitting even at higher resolutions. Future work is needed to study the potential for tropical cyclogenesis in higher-resolution models of tidally locked terrestrial planets, including local convection-resolving large eddy simulations \citep{Lefevre:2021aa}, global convection-resolving simulations \citep{Yang:2023aa}, or global simulations with {nested grids} \citep{Sergeev:2020aa}. Such work would facilitate comparison between coarse global tropical cyclone-permitting simulations and tropical cyclone-resolving and convection-permitting simulations in order to determine the extent to which the methods used in this work are broadly applicable to study tropical cyclogenesis on tidally locked terrestrial exoplanets.

\section{Conclusions}
\label{sec:conc}
In this work, we conducted GCM simulations of tidally locked terrestrial exoplanets with a thermodynamically active slab ocean and sufficient spatial resolution to permit tropical cyclogenesis. We conducted simulations with varying rotation periods of 4-16 days, and found that tropical cyclones occur in each simulation. We further find that the number and strength of tropical cyclones in each simulation agrees with that predicted from Earth-based environmental favorability metrics.  Our key conclusions are as follows:
\begin{enumerate}[topsep=12pt,itemsep=12pt,partopsep=1ex,parsep=1ex]
    \item Tropical cyclones occur in our ExoCAM GCM simulations of tidally locked terrestrial exoplanets orbiting late-type M dwarf stars that have a sufficient horizontal resolution to permit tropical cyclogenesis. Notably, these GCM simulations include a slab ocean, while previous work by \cite{YanYang:2020aa} used a prescribed sea surface temperature distribution. Thus, we extend the finding of \cite{YanYang:2020aa} that tropical cyclones occur in fixed sea surface temperature GCMs of tidally locked terrestrial exoplanets to simulations with thermodynamically consistent sea surface temperatures. 
    
    \item All tropical cyclones in our GCM experiments occur on the dayside of the planet, near the substellar point. The locations of the simulated tropical cyclones broadly match with environmental favorability metrics based on Earth and resulting predictions from the low-resolution simulations of \cite{Komacek:2020ad}, especially the combination of the ventilation-reduced maximum potential intensity and the absolute vorticity. Notably, we do not find warm-core tropical cyclones near the poles and on the nightside, but they did form in the fixed sea surface temperature simulations of \citep{YanYang:2020aa}. The lack of tropical cyclones on the cooler poles and nightside is in agreement with expectations based on the ventilation index, which is small or undefined in these regions due to a thermal inversion in the boundary layer.  
    
    \item The dependence of the number and strength of tropical cyclones on rotation period matches well with the expectations from the low-resolution GCMs studied in \cite{Bin:2018aa} and \cite{Komacek:2020ad}. We find a peak in both the occurrence of tropical cyclones and their mean wind speeds in the case with an intermediate rotation period of 8 days. This local maximum in tropical cyclone intensity and number agrees with expectations from the coherent alignment of regions of high values of ventilation-reduced maximum potential intensity and absolute vorticity. The case with a rotation period of 4 days has an intermediate number of storms compared to the two slower-rotating cases, and storms in this fast-rotating case are associated with the highest average maximum precipitation.
    We further find no tropical cyclones with hurricane-force winds in the slow-rotating case with a rotation period of 16 days, as expected from the combination of a small absolute vorticity and weak maximum potential intensity. 
\end{enumerate}

\acknowledgments
 We thank Eric Wolf for making ExoCAM freely available to the community and continuing its development. {We thank the anonymous referee for their insightful comments and suggestions, which greatly improved the manuscript.} We thank Tobi Hammond and Julianna Heptinstall for helpful feedback on this work. We acknowledge generous support from the Heising-Simons Foundation as part of the 51 Pegasi b Fellowship Enhancements. CMS and TDK are grateful to have received support from the 2023 Burgers Program Summer Research Scholarships. We acknowledge high-performance computing support from Cheyenne (\url{doi:10.5065/D6RX99HX}) provided by NCAR's Computational and Information Systems Laboratory, sponsored by the National Science Foundation under NSF AGS grant 2209052. We acknowledge the University of Maryland supercomputing resources (\url{http://hpcc.umd.edu}) made available for conducting the research reported in this paper. 

\bibliography{References_terrestrial_hurricane,refs_CHAVAS}

\end{document}